\documentclass[superscriptaddress,twocolumn,showpacs,pra]{revtex4}

\usepackage[usenames]{color}
\usepackage{amsmath,amssymb}
\usepackage{graphics}
\usepackage{epsfig}

\begin{document}

\title{Efimov-van-der-Waals universality for ultracold atoms with positive scattering lengths}

\author{Paul M. A. Mestrom}
\affiliation{Eindhoven University of Technology, P. O. Box 513, 5600 MB Eindhoven, The Netherlands}

\author{Jia Wang}
\affiliation{Centre for Quantum and Optical Science, Swinburne University of Technology, Melbourne, Australia}

\author{Chris H. Greene}
\affiliation{Department of Physics and Astronomy, Purdue University, West Lafayette, Indiana 47907, USA}

\author{Jos\'e P. D'Incao}
\affiliation{JILA, University of Colorado and NIST, Boulder, Colorado 80309, USA}
\affiliation{Department of Physics, University of Colorado, Boulder, Colorado 80309, USA}

\begin{abstract}
{We study the universality of the three-body parameters for systems relevant for ultracold quantum gases with 
positive $s$-wave two-body scattering lengths.} 
Our results account for finite-range effects and their universality is tested
by changing the number of deeply bound diatomic states supported by our interaction model. We find
that the physics controlling the values of the three-body parameters associated with the ground and excited Efimov states
is constrained by a variational principle and can be strongly affected by $d$-wave interactions that prevent both trimer states from 
merging into the atom-dimer continuum. Our results enable comparisons to current experimental data and they suggest tests 
of universality for atomic systems with {positive scattering lengths}.
\end{abstract}

\maketitle 

\section{Introduction}

The recent theoretical and experimental progress in the exploration of ultracold quantum gases in the strongly interacting 
regime have largely established the relevance of the three-body Efimov physics \cite{Efimov1970,braaten2006pr,wang2013aamop}
for the understanding of both dynamics and stability of such systems 
\cite{fletcher2013prl,rem2013prl,makotyn2014np,fletcher2016arxiv,sykes2014pra,laurent2014prl,smith2014prl,piatecki2014nc,barth2015pra}.
The control of interatomic interactions through magnetic-field dependent Feshbach resonances \cite{chin2010rmp} allows 
for dramatic changes in the $s$-wave two-body scattering length, $a$, making it possible to tune systems to the vastly
different collective {(mean-field)} regimes of attractive, $a<0$, and repulsive, $a>0$ interactions. In the regime of strong interactions, 
$|a|/r_{\rm vdW}\gg1$, where $r_{\rm vdW}$ is the van der Waals length \cite{chin2010rmp}, the Efimov effect is manifested through the
appearance of an infinite series of three-body states that can lead to scattering resonances and 
interference effects accessible to experiments \cite{braaten2006pr,wang2013aamop}. Such dramatic few-body phenomena open up the possibility to explore
new quantum regimes in ultracold gases. One of the striking signatures of the Efimov effect is the geometric scaling
of the system for many trimer properties, which interrelates all the three-body observables via the geometric factor 
$e^{\pi/s_0}$, where $s_0\approx1.00624$ for identical bosons. As a result, if universal scaling holds, the determination of a single observable
---the {\em three-body parameter}--- would allow derivation of all properties of the system.
However, since the early days of Efimov's original prediction it was largely accepted that this three-body parameter
would be different for every system. Nevertheless, a few years ago, as experiments in ultracold gases evolved, it became clear
that this concept needed reassessment.

The turnaround came from the experimental observations in $^{133}$Cs \cite{berninger2011prl} showing that the three-body 
parameter $a_-$, associated with the value of $a<0$ at which the first Efimov state merges with the three-body continuum, 
were the same (within a 15\% margin) for different resonances in $^{133}$Cs. Moreover, if the results were recast in terms of $r_{\rm vdW}$, the 
observations in every other available atomic species also led to similar results, $a_-/r_{\rm vdW}\approx-10$ (see Ref.~\cite{wang2013aamop} 
for a summary of such experimental findings). Theoretical works then successfully confirmed and interpreted the universality
of the $a_-$ parameter \cite{wang2012prl,dincao2013fbs,wang2012prlb,schmidt2012epjb,naidon2014pra,naidon2014prl,blume2015fbs} 
and consolidated a new universal picture for Efimov physics in atomic systems dominated by van der Waals forces.

This {paper} assesses the universality of the three-body parameter in the yet unexplored regime of 
{positive} scattering lengths, 
$a>0$. The available experimental data for Efimov features within this regime 
is relatively sparse and, consequently, does not clearly display the same degree of universality found for $a<0$. 
{Although not explicitly demonstrated here, our present theoretical study 
shows that universality for $a>0$ persists and is rooted in the same 
suppression of the probability of finding particles at short distances previously found 
for $a<0$ \cite{wang2012prl,dincao2013fbs,wang2012prlb}.}
The observables we analyze are related to the value of $a$ at which an Efimov state
intersects the atom-dimer threshold, $a_*$, thus causing a resonance in atom-dimer collisions 
\cite{dincao2005prl,braaten2007pra}, and the value $a_+$ at which a minimum in three-body recombination 
occurs as a result of a destructive interference between the relevant 
collision pathways \cite{nielsen1999prl,esry1999prl,bedaque2000prl,dincao2005prl}.
One important feature that can help to interpret our computed values for $a_*$ and $a_+$ associated with 
the ground Efimov state is the existence of a variational principle \cite{bruch1973prl,lee2007pra} 
{that constrains its energy to always lie below a certain value lower than the dimer energy, thus preventing the
trimer to cross the atom-dimer threshold.}
This has a direct impact on both the lowest atom-dimer resonance and on interference phenomena, even when, as we show here,
the conditions for the validity of that variational principle are not strictly satisfied.  
Moreover, our analysis indicates that the presence of strong $d$-wave interactions 
\cite{gao2000pra,wang2012pra}{, and/or possibly some other finite-range effects,}
also prevents the first excited Efimov state from merging with the dimer threshold, although it still produces a 
resonance feature in atom-dimer observables and a recombination minimum for small $a$. 

\section{Brief theoretical background}

Here we use the adiabatic hyperspherical representation which offers a simple and conceptually clear 
description of few-body systems while still accurately determining their properties \cite{wang2013aamop}. Within this representation, 
after solving for the hyperangular internal motion ---which includes all interparticle interactions--- three-body observables 
can be obtained by solving the hyperradial Schr\"odinger equation  \cite{wang2011pra}
\begin{multline}
\left[-\frac{\hbar^2}{2\mu}\frac{d^2}{dR^2}+W_{\nu}(R)\right]F_{\nu}(R)
 \\+\sum_{\nu'\neq\nu} W_{\nu\nu'}(R) F_{\nu'}(R)=E F_\nu(R).\label{RadialEq}
 \end{multline}
where the hyperradius $R$ describes the overall size of the system, $\mu=m/\sqrt{3}$ is the three-body
reduced mass and $\nu$ is an index including all necessary quantum numbers to characterize each channel. 
Equation~(\ref{RadialEq}) describes the radial motion governed by the effective hyperspherical
potentials $W_\nu$ and non-adiabatic couplings $W_{\nu\nu'}$, which determine all
bound and scattering properties of the system.
In the present study, each pair of particles interacts via a Lennard-Jones
potential 
\begin{equation}
v_{LJ}(r) = - \frac{C_6}{r^6}\left(1-\frac{\lambda^6}{r^6}\right),
\label{vLJ}
\end{equation}
where $\lambda$ is adjusted to give the desired value of $a$ and $C_6$ is the usual dispersion coefficient. Note that our
calculations use van der Waals units (with energy and length units of $E_{\rm vdW}=\hbar^2/mr_{\rm vdW}^2$
and $r_{\rm vdW}$) such that the specification of the value of $C_6$ is unnecessary. 
Our present study is centered around the first three poles of $a$, which occur at the values denoted $\lambda=\lambda^*_{1}$,
$\lambda^*_{2}$ and $\lambda^*_{3}$. One important point to keep in mind is that near $\lambda^*_{1}$ 
there can exist only a single two-body $s$-wave state, whereas near $\lambda^*_{2}$ and $\lambda^*_{3}$ multiple deeply bound states 
exist (4 and 9, respectively), owing to the presence of higher partial wave dimers.

\section{Results}

\begin{figure}[hbtp]
\includegraphics[width=3.4in]{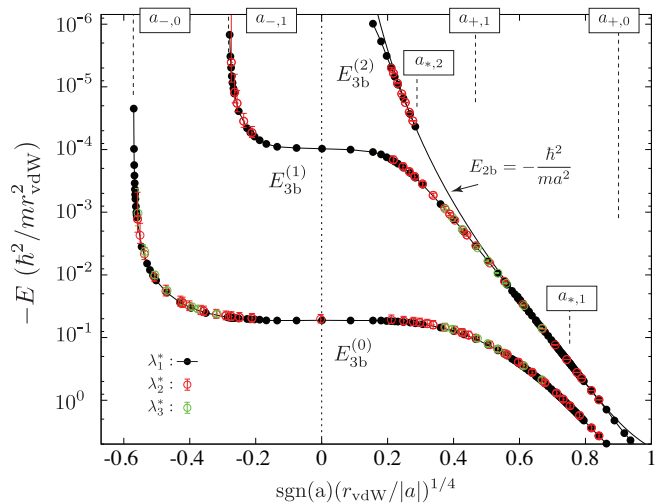}
\caption{Energy of Efimov states calculated near the first three poles of $a$, $\lambda=\lambda^*_1$, $\lambda^*_2$
and $\lambda^*_3$, in our model potential in Eq.~(\ref{vLJ}). Near $\lambda^*_1$, Efimov states (black filled circles) are true bound states
while near the $\lambda^*_2$ and $\lambda^*_3$ (red and green open circles, respectively) Efimov
states are resonant states with the corresponding widths indicated as the error bars. Approximated values for $a_-$, $a_*$, and $a_+$ are also indicated.}
\label{EfimovE}
\end{figure}

Figure \ref{EfimovE} shows the energies of the lowest three Efimov states, $E_{\rm 3b}$, for values of 
$a$ near the three poles considered ($\lambda^*_{1}$, $\lambda^*_{2}$ and $\lambda^*_{3}$),
offering a global view of the degree of the universality of our results. 
Near $\lambda^*_{1}$, Efimov states (black filled circles) are true bound states
while near $\lambda^*_{2}$ and $\lambda^*_{3}$ (red and green open circles, respectively) Efimov
states are resonant states whose (presumably nonuniversal) widths have been calculated using the Ref.~\cite{nielsen2002pra} procedure,
indicated in Fig.~\ref{EfimovE} as the error bars. 
The atom-dimer threshold, defined by the dimer energy, $E_{\rm 2b}=-\hbar^2/ma^2$ ($a\gg r_{\rm vdW}$), is also shown 
(solid line).
In Fig.~\ref{EfimovE} the ground Efimov state does not ``cross" or intersect the atom-dimer threshold, as expected
from the variational principle in Refs.~\cite{bruch1973prl,lee2007pra}, which state that $E_{\rm 3b}<3E_{\rm 2b}$ .
In principle, this variational constraint applies only to bound states, i.e., only for Efimov states near $\lambda^*_{1}$,
however, our calculations for the energies of Efimov resonances near $\lambda^*_2$ and $\lambda^*_3$ also
follow the same non-crossing rule.
Evidently, this effect strongly modifies the expected 
universality predicted by zero-range models since it prevents an atom-dimer resonance and can also modify the minima in recombination associated 
with the ground Efimov state. 
Table \ref{Tab3BP} summarizes our computed values of the three-body parameters
---see also Fig.~\ref{EfimovE} for their approximate location. [The values for $a_-$ were previously determined in
Ref.~\cite{wang2012prl} (and in unpublished work from that study).] The additional index on the $a_-$, $a_+$, and $a_*$ parameters 
indicates their Efimov family parentage. The physics involved and caveats on the determination of these
three-body parameters are given below.

\begin{table}[h]
\caption{Values for the three-body parameters $a_-$, $a_*$ and $a_+$ for the lowest two Efimov scattering features in 
recombination and atom-dimer collisions, near the lowest three poles in the scattering length. For $a_{+,1}$ we also show
its dependence on the temperature by $\langle K_3\rangle$ (see text) at values of $k_BT/E_{\rm vdW}$ (indicated in square brackets)
listed in the last three columns below. {In the bottom part of the table we list the universal ratios $\theta_{ij}^{\alpha\beta}$ 
[see Eq.~(\ref{uratios})] resulting from the average value of the three-body parameters (see text for the comparison with the zero-range results).}}
\begin{ruledtabular}
\begin{tabular}{ccccccc}
        & $a_{-,i}/r_{\rm vdW}$ & $a_{*,i}/r_{\rm vdW}$ & $a_{+,i}/r_{\rm vdW}$ & \multicolumn{3}{c}{$\langle a_{+,1}\rangle/r_{\rm vdW}$} \\
Pole & ($i=0,1$) & ($i=1,2$) & ($i=0,1$) & [$10^{\mbox{-}4}$] & [$3\times10^{\mbox{-}4}$] & [$10^{\mbox{-}3}]$ \\
 \hline
 $\lambda^*_1$ & -$9.60$, -$161$  & $3.41$, $157$ & $1.41$, $27.2$& $28.0$ & $29.1$ & $32.1$ \\
 $\lambda^*_2$ & -$9.74$, -$164$  & $3.26$, $160$ & $1.41$, $27.9$ & $28.7$ & $30.7$ & $34.8$ \\
 $\lambda^*_3$ & -$9.96$, ---          & $3.33$, $160$ & $1.41$, $28.0$ & --- & --- & --- \\[0.05in]
Avg. & -$9.77$, -$163$ & $3.33$, $159$ & $1.41$, $27.7$ & $28.4$ & $29.9$ & $33.5$ \\ [0.05in]
\hline
$(i,j)$ & $(0,0)$ & $(0,1)$ & $(1,0)$ & $(1,1)$ & $(2,0)$ & $(2,1)$ \\
 \hline
 $\theta_{ij}^{+-}$ 
 & 0.143 & 0.195 & 0.125 & 0.170 & --- & --- \\
  $\theta_{ij}^{*-}$ 
 & --- & --- & 0.015 & 0.020 & 0.032 & 0.043 \\
   $\theta_{ij}^{*+}$ 
 & --- & --- & 0.105 & 0.120 & 0.220 & 0.253 \\
\end{tabular}
\end{ruledtabular}
\label{Tab3BP}
\end{table}

\begin{figure}[hbtp]
\includegraphics[width=3.4in]{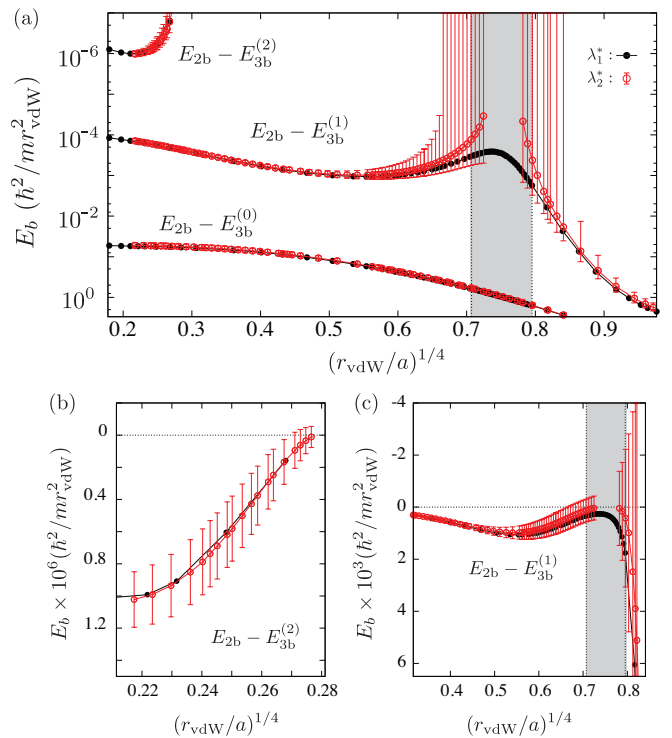}
\caption{(a) Binding energies, $E_b=E_{\rm 2b}-E_{\rm 3b}$, of Efimov states near $\lambda^*_1$ and $\lambda^*_2$ 
(black filled and red open circles, respectively) showing that both ground and first excited Efimov states fail to merge 
into the atom-dimer threshold (see text). {In (b) and (c) we show a blow up of (a) near the second and first
excited Efimov states, respectively.}}
\label{EfimovEb}
\end{figure}

Closer inspection of Fig.~\ref{EfimovE} reveals that the first excited Efimov state also fails to intersect with the dimer
threshold. This is clearly shown in Fig.~\ref{EfimovEb} for the binding energy of the Efimov states, $E_b=E_{\rm 2b}-E_{\rm 3b}$.
Near $\lambda^*_1$ (black filled circles) the non-crossing of the first excited state is evident within the shaded region in Fig.~\ref{EfimovEb}. 
Near $\lambda^*_2$ (red open circles), the qualitative behavior is the same, however: As the energy of the
Efimov state approaches the threshold its width increases to the point in which it exceeds the value of its binding energy  
---therefore, losing some its ``bound" state character--- and eventually ``dissolving'' into the atom-dimer continuum 
(see shaded region in Fig.~\ref{EfimovEb}).
Passing this point, as $a$ decreases further, the state recovers its bound character. 
{Our physical interpretation of the non-crossing of the first excited Efimov state \cite{comment0} is that it results 
from the existence of strong $d$-wave interactions near $a/r_{\rm vdW}=1$ \cite{gao2000pra,wang2012pra}. 
Within our theoretical model, since $s$- and $d$-wave interactions can not be separated, a more clear physical 
picture of the non-crossing of the first excited Efimov state still remains, leaving even the possibility of that being a generalization
of the same variational principle \cite{bruch1973prl,lee2007pra} which prevents the ground state to unbind.}
Figure \ref{EfimovEb} shows that only the second excited Efimov state displays the expected intersection with the atom-dimer threshold.

\begin{figure}[hbtp]
\includegraphics[width=3.4in]{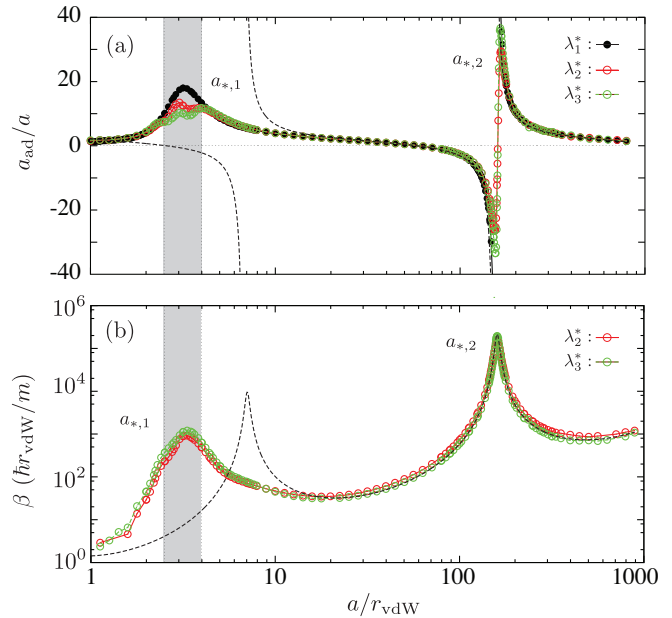}
\caption{(a) The atom-dimer scattering length, $a_{\rm ad}$, and (b) corresponding loss rate, $\beta$, 
displaying resonant behavior due to Efimov resonances associated with the first and second excited Efimov states.
The values of the three-body parameters $a_{*,1}$ and $a_{*,2}$ are indicated in the figure. The dashed curve gives the 
analytical zero range results from Ref.~\cite{braaten2007pra}.}
\label{ADRates}
\end{figure}

Evidently, the effects analyzed above have an important impact on the determination
of the three-body parameter $a_*$. This is achieved here by directly calculating the corresponding 
atom-dimer scattering properties. 
Of particular importance for ultracold experiments is the atom-dimer scattering length 
$a_{\rm ad}$ and the atom-dimer loss rate $\beta$ \cite{dincao2008pra}.
Figure~\ref{ADRates} shows our calculated values for these quantities. 
In Fig.~\ref{ADRates} (a), around the shaded region (corresponding to the same shaded region in Fig.~\ref{EfimovEb})
$a_{\rm ad}$ is enhanced, however, remaining always positive and consistent with the failure of the
first excited Efimov state in Fig.~\ref{EfimovEb} to become unbound. (Note that in this regime $a_{\rm ad}$ for $\lambda^*_2$ and $\lambda^*_3$ 
displays a more complicated dependence on $a$ due to the presence of strong couplings to nearby 
three-body channels.)
For larger $a$, $a_{\rm ad}$ is now enhanced and changes sign, implying that the second 
excited Efimov state intersects with the dimer energy (see Fig.~\ref{EfimovEb}). 
Note that here, $a_{\rm ad}$ for $\lambda^*_2$ and $\lambda^*_3$ 
does not actually diverge due to the presence of inelastic processes \cite{hutson2007njp}.
Figure~\ref{ADRates}(b) shows the corresponding atom-dimer loss rates, which display the resonant behavior associated with
the first and second excited Efimov states. 
Even though the first excited Efimov state does not become unbound, it approaches the atom-dimer threshold close enough 
to produce a clear enhancement in the atom-dimer loss rate. 
We define $a_{*,1}$ and $a_{*,2}$ as the value of $a$ where 
$\beta$ is maximum [see Fig.~\ref{ADRates}~(b)], except for our calculations near the first pole, where no 
losses occur ($\beta=0$). 
In this case $a_{*,1}$ and $a_{*,2}$ were determined from the maximum value of $a_{\rm ad}$ [see 
Fig.~\ref{ADRates}(a)]. Numerical values are listed in Table~\ref{Tab3BP}. In order to contrast 
our numerical results with the universal predictions (based on two-body contact interaction models),
we also display in Fig.~\ref{ADRates} (dashed lines) the expected behavior for $a_{\rm ad}$ and $\beta$ 
from Ref.~\cite{braaten2007pra}. 
{For the zero-range, universal, model of Ref.~\cite{braaten2007pra} we used
the averaged value for $a_{*,2}$ from Table~\ref{Tab3BP} as the three-body parameter, and set the inelasticity parameter 
$\eta=0$ in Fig.~\ref{ADRates}(a) and $\eta=0.03$ for Fig.~\ref{ADRates}(b), in order to better fit the data for $\lambda^*_3$.}
Although the agreement is very good for large $a$, near 
$a_{*,1}$ not only finite range corrections become more important but also the fact that the first excited Efimov 
state fails to intersect with the atom-dimer threshold, imply strong deviations 
between universal zero-range theory and our results.

\begin{figure}[hbtp]
\includegraphics[width=3.4in]{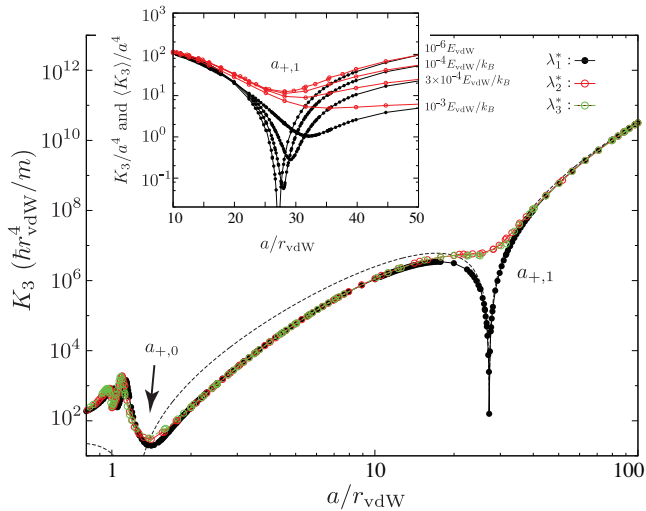}
\caption{Three-body recombination, $K_3$, displaying interference minima
associated with the ground and first excited Efimov states. Values of the three-body parameters $a_{+,0}$
and $a_{+,1}$ are indicated in the figure. The dashed curve gives the analytical zero range 
result in the absence of deeply bound dimers \cite{braaten2006pr,comment1}.
{\em Inset}:  Thermally averaged recombination rate, $\langle K_3\rangle$, 
illustrating the temperature dependence of $\langle a_{+,1}\rangle$.}
\label{K3Rates}
\end{figure}

Finally, we have also calculated the three-body 
recombination rate, $K_3$, in the lowest three-body angular 
momentum ($J=0$) \cite{suno2002pra,dincao2004prl} 
to determine the values of the three-body parameter $a_{+}$. 
Figure \ref{K3Rates} shows our results
for $K_3$ in the zero-energy limit ($E=10^{-6}E_{\rm vdW}$) clearly displaying two minima, whose locations 
are identified as the values for $a_{+,0}$ and $a_{+,1}$ listed
in Table \ref{Tab3BP}. Our numerical results obtained near $\lambda^*_1$ are compared
with the analytical results in the absence of deeply bound dimers \cite{braaten2006pr,comment1} (dashed line). 
For large $a$ our results agree well with the analytical ones while strong deviations can be observed 
for small $a$. In particular, one can see that the predicted
minimum in recombination near $a/r_{\rm vdW}=1$ is strongly affected by finite-range effects. 
We trace such effects to the
presence of strong $d$-wave interactions \cite{comment0}. In fact, near $a/r_{\rm vdW}=1$ our results display an enhancement due to
a universal three-body resonance with strong $d$-wave character \cite{wang2012pra}. Therefore, 
our result for $a_{+,0}$ is a balance between universal $s$- and $d$-wave physics \cite{comment0}. The inset of Fig.~\ref{K3Rates}
shows the temperature dependence of $K_3$ obtained by calculating the thermally averaged recombination rate $\langle K_3\rangle$
\cite{dincao2004prl}, which illustrates the temperature dependence of $\langle a_{+,1}\rangle$ in the regime
relevant for experiments ---see also the values listed in Table \ref{Tab3BP}. 
{In principle, at finite temperatures one would also need to include higher partial-waves contributions
to recombination. For identical bosons, however, the next leading contribution is for $J=2$ and scale with the
temperature and scattering length as $T^2a^8$ \cite{dincao2004prl}. In that case, for the temperatures we explore 
in Fig.~\ref{K3Rates} and values for $\langle a_{+,1}\rangle$ listed Table \ref{Tab3BP}, such effects are likely 
to be small, except perhaps for our largest temperature, where $ka_{+,1}\approx0.88$ (see also the analysis 
in Ref.~\cite{dincao2004prl}).}

Our results for the three-body parameters ---summarized in Table \ref{Tab3BP}--- clearly
show universal behavior 
{(with deviations between themselves within a few percent)}
and should be applied for atomic species 
with isolated broad Feshbach resonances. 
We also used our results in Table~\ref{Tab3BP} to determine other universal properties ---for instance, the 
ratios $a_+/a_-$, $a_*/a_-$, and $a_*/a_+$--- and compare with those resulting from zero-range models 
\cite{braaten2006pr,gogolin2008prl}.
{
For that we define the ratio between different three-body parameters as
\begin{eqnarray}
a_{\alpha,i}/a_{\beta,j}=\theta_{ij}^{\alpha\beta}(e^{\pi/s_0})^{i-j}, \label{uratios}
\end{eqnarray}
where $\alpha$ and $\beta$ can assume the values ``$-$", ``$+$" and ``$*$", while $i$ and $j$ run over the index labeling 
the Efimov state. Within the zero-range model $\theta$ is a universal number and does not depend on $i$ and $j$: 
$\theta_{ij}^{+-}\approx0.210$; $\theta_{ij}^{*-}\approx0.047$; $\theta_{ij}^{*+}\approx0.224$ \cite{braaten2006pr,gogolin2008prl}. 
Comparing those with the ones shown in the bottom part of Table~\ref{Tab3BP} (calculated using the averaged values for $a_-$, $a_+$ 
and $a^*$) we have found substantial deviations, most likely due to finite-range effects and the absence of $d$-wave interactions 
in the zero-range model. 
Moreover, the values for the geometric scaling factors obtained from our calculations, $a_{-,1}/a_{-,0}\approx16.7$, $a_{+,1}/a_{+,0}\approx19.7$, and 
$a_{*,2}/a_{*,1}\approx47.8$, also display strong deviations from the universal value $e^{\pi/s_0}\approx22.69$.
We note that the results for the geometric scaling factor for $a_-$ obtained in Refs.~\cite{deltuva2012pra,schmidt2012epjb} 
are consistent to ours but the corresponding results for $a_*$ from \cite{schmidt2012epjb} are not. A comparison with results 
originated from models which include finite-range corrections \cite{schmidt2012epjb,deltuva2012pra,kievsky2013pra,ji2012ap,platter2009pra,ji2015pra}
needs to be made carefully to ensure that the interaction parameters are the same. This, however, is beyond the scope of the present study.
A more direct comparison, however, can be made with the work in Ref.~\cite{li2016pra}, where a model similar to ours, however,
considering only $s$-wave interactions{, is used. The calculations of Ref.~\cite{li2016pra} involve a separable approximation of a hard-core-type van der Waals potential as two-body interaction potential.} The comparison between our results and the ones from Ref.~\cite{li2016pra} thus
provides a sense of how important $d$-wave interactions might be. In Table~\ref{TabTheoComp} we list our average results, marked by LJ,
(see Table~\ref{Tab3BP}) and the corresponding average results from Ref.~\cite{li2016pra}, marked by LJ$^s$. 
In Table~\ref{TabTheoComp} we also list the value of $\kappa_0=(mE_{\rm 3b}^{(0)}/\hbar^2)^{1/2}$ obtained from 
our calculations for $\lambda^*_1$ and the corresponding averaged result from Ref.~\cite{li2016pra}. 
The agreement is generally good for all cases (the relative errors are indicated in Table~\ref{TabTheoComp} between square brackets), 
with the exception for the value of $a_{*,1}$, most likely because the non-crossing of the first excited Efimov state is absent in the 
model of Ref.~\cite{li2016pra}, clearly indicating a strong effect due to $d$-wave interactions. We note, however, that the agreement 
for the geometric factors $a_{-,1}/a_{-,0}$ and $a_{+,1}/a_{+,0}$ are generally better than the absolute values of the three-body
parameters. This indicates that the effect of the $d$-wave interactions in such parameters is mainly to introduce a shift:
\begin{align}
{a}_{x}\rightarrow{a_x}e^{-\phi_d/s_0},\label{Shift3BP}
\end{align} 
or, equivalently, a change in the three-body phase: $s_0\ln({a}/{a_x})\rightarrow s_0\ln({a}/{a_x})+\phi_d$.
Indeed, forcing our value of $a_{+,0}$ to reproduce the one from Ref.~\cite{li2016pra}, we obtain $\phi_d\approx-0.146$
and the resulting rescaled three-body parameters, marked by LJ$^*$ in Table~\ref{TabTheoComp} now agree
much better, evidently, with the exception of $a_{*,1}$. The above rescaling process, therefore, can be seen as an 
attempt to subtract-off $d$-wave effects from our calculations, although a more rigorous study that can provide a more
quantitative analysis of such effects still needs to be performed.}

\begin{table}[h]
\caption{{Comparison between the average results for the three-body parameters in Table~\ref{Tab3BP}, marked here by LJ,
 and the corresponding average results from Ref.~\cite{li2016pra}, marked by LJ$^s$. 
The Table also lists the value of $\kappa_0=(mE_{\rm 3b}^{(0)}/\hbar^2)^{1/2}$ obtained from 
our calculations for $\lambda^*_1$ and the corresponding averaged result from Ref.~\cite{li2016pra}. 
The corresponding relative errors are indicated between square brackets.}}
\begin{ruledtabular}
\begin{tabular}{lcccccc}
 & ${a}_{-,i}$ & ${a}_{-,1}/{a}_{-,0}$ & $\kappa_0$ & ${a}_{+,i}$ & ${a}_{+,1}/{a}_{+,0}$ & ${a}_{*,1}$ \\
  & $(i=0,1)$ &  &  & $(i=0,1)$ &  &  \\
\hline
LJ                   & -9.77,-163  & 16.7   & 0.230   & 1.41,27.7   & 19.7  & 3.33   \\
LJ$^s$  & -10.7,-187  & 17.5   & 0.193   & 1.63,33.5   & 20.6  & 5.49   \\
                       & [0.10,0.15] & [0.05] & [0.16] & [0.16,0.21] & [0.05] & [0.65] \\
LJ$^*$ & -11.3,-188  & 16.7   & 0.199   & 1.63,32.0   & 19.7   & 3.85   \\
                       & [0.05,0.01] & [0.05] & [0.03] & [0.00,0.05] & [0.05] & [0.43] \\
\end{tabular}
\end{ruledtabular}
\label{TabTheoComp}
\end{table}

\begin{table}[h]
\caption{Experimental values for the three-body parameters $a_+$ and $a_*$. The table displays our assignment of the parameters
by indicating the value of $i$ for $a_{+,i}$ and $a_{*,i}$ for each case. We also list the values for $a/a_c$ \cite{dincao2004prl} 
characterizing the degree of thermal effects in the experimental data.}
\begin{ruledtabular}
\begin{tabular}{ccccccc}
Atom & $a_+/a_c$ & $i$ & $a_{+,i}/r_{\rm vdW}$ & $a_{*}/a_c$ & $i$ & $a_{*,i}/r_{\rm vdW}$ \\
\hline
$^{133}$Cs 
& 0.08 & 0 & 2.1(0.1) \cite{Kraemer2006} & 0.13 & 1 & 4.2(0.1) \cite{Knoop2009,Zenesini2014}\\
 & 0.03 &0 & 2.7(0.3)\cite{berninger2011prl} &0.24 & 1 & 6.5(0.3) \cite{Zenesini2014}\\
 & ---    & 0 & 2.5(0.4) \cite{Ferlaino2011} & & \\
$^7$Li 
& 0.02 & 0 & 2.7(0.1)\cite{Pollack2009,Dyke2013} & 0.09 & 1 &13.0(0.6) \cite{Pollack2009,Dyke2013}$^\dagger$ \\
& 0.29 & 1 & 44(3) \cite{Pollack2009,Dyke2013} & 0.04 & 1 & 5.5 \cite{Machtey2012a} \\
& 0.32 & 1 & 35(4) \cite{Gross2009,Gross2011} & 0.05 & 1 & 6.0(0.1) \cite{Machtey2012a}$^\dagger$ \\
& 0.34 & 1 & 39(2) \cite{Gross2010,Gross2011} & & \\
$^{39}$K 
& 0.03 & 0 & 3.5(0.1) \cite{Zaccanti2009} & 0.01 & 0 & 0.5(0.2) \cite{Zaccanti2009}$^\dagger$ \\
& 0.76 & 1 & 88(14) \cite{Zaccanti2009} & 0.12 & 1 & 14.4(0.6) \cite{Zaccanti2009}$^\dagger$ \\
$^{6}$Li 
  & & & & 0.01 & 1& 2.9 \cite{Lompe2010a}
\end{tabular}
\end{ruledtabular}
\label{TabExp}
\end{table}

We now analyze the currently available experimental data for $a_+$ and $a_*$ listed (and assigned) in Table \ref{TabExp}. 
{As one can see from Table \ref{TabExp}, the values listed for $a_{+,0}$ and $a_{+,1}$ are qualitatively 
consistent among themselves, with the exception of the data for $^{39}$K \cite{Zaccanti2009} ---a new analysis
presented in Ref.~\cite{roy2013prl} suggests that this data might be subject of a new calibration.
Although the values for $a_{+,1}$ in Table \ref{TabExp} are likely to suffer from thermal
effects (the condition $|a|\ll a_c = {\hbar}/{\sqrt{m k_B T}}$ \cite{dincao2004prl} ensuring the absence of thermal effects 
is not strictly satisfied), our finite temperature calculations covering the range of temperatures relevant for the experiments 
(see Table~\ref{Tab3BP}) indicate that thermal effects might lead to no more than a 10\% variation from the zero temperature 
result. 
We also note that for $^7$Li and $^{39}$K the resonances are substantially less broad than the ones for $^{133}$Cs 
(see Ref.~\cite{chin2010rmp}), thus opening up the possibility of finite-width effects as responsible for the deviations 
among the experimental data in Table~\ref{TabExp}.
In comparison to the values for $a_+$, the results for $a_{*,1}$ listed in Table \ref{TabExp} 
display a much stronger deviation among themselves. 
A more careful analysis, therefore, is necessary in order to understand some of the possible factors 
affecting such observations. For instance, the value for $a_{*,1}$ for $^{133}$Cs from Ref.~\cite{Zenesini2014},
as well as the results for $^{7}$Li from Ref.~\cite{Machtey2012a}, were obtained using a Feshbach resonance
that is not well separated from another nearby resonance, possibly affecting the observed value for $a_{*,1}$.
Most of the results marked in Table~\ref{TabExp} by ``$^\dagger$'' present the largest variations compared with the total
averaged result for $a_{*,1}$ ($\approx6.63r_{\rm vdW}$).
They were, however, obtained based on the assumption that atom-dimer resonances can be observed in atomic 
samples by means of an avalanche mechanism \cite{Zaccanti2009}.
Although modifications on the description of such mechanism can lead to more reasonable results \cite{Machtey2012a,Machtey2012b},
this hypothesis is currently considered questionable \cite{Langmack2012,Zenesini2014,Hu2014}.}

{Therefore, accordingly to our analysis above, in order to properly compare the experimental data to 
theoretical predictions, we excluded the data from $^{39}$K \cite{Zaccanti2009} and those marked by ``$^\dagger$'' 
in Table~\ref{TabExp}. From the remaining experimental data, we determine an average value and corresponding relative 
error as listed in Table~\ref{TabExpComp} (the relative errors are indicated between square brackets).
Using the zero-range (ZR) universal relations derived in Refs.~\cite{braaten2006pr,gogolin2008prl}
we determined the values for $a_{+,0}$, $a_{+,1}$ and $a_{*,1}$, using the average value for $a_{-,0}$ in Table~\ref{Tab3BP},
and list these in Table ~\ref{TabExpComp}, along with our corresponding averaged results (LJ) from Table~\ref{Tab3BP}.
As one can see, the zero-range results for $a_{+,0}$ and $a_{+,1}$ perform better than our results when compared to
the experimental data, while our result for  $a_{*,1}$ outperforms the zero-range result. In fact, within the zero-range model the atom-dimer 
resonance associated to $a_{*,1}$ originates from an actual crossing between the first excited Efimov state while in 
our model it does not (see Fig.~\ref{EfimovEb} and corresponding discussion in the text).
We note, however, that our results for $a_{+,1}/a_{+,0}$ better reproduces the value from the experimental data,
indicating that a shift on the position of the three-body parameters for $a>0$, in the same spirit than the one obtained
from Eq.~(\ref{Shift3BP}), can improve the comparisons. In fact, as shown in Table~\ref{TabExpComp}, 
using the results (LJ$^*$) listed in Table~\ref{TabTheoComp}, an overall improvement can be observed.
If we now rescale our results (LJ) to reproduce the experimental value for $a_{+,0}$, within a relative error of 0.10, 
our new rescaled results (LJ$^*_{\rm E}$) now display a much better overall comparison to the experimental data. 
Although there is no clear reason why such scaling should be allowed, the above analysis clearly indicates that our 
numerical results might suffer from finite-range effects, whether originated from the $s$- and $d$-wave mixing or the finite-width character
produced by real interatomic interactions. }

\begin{table}[h]
\caption{{Comparison between the values for the three-body parameters from
different theories and the average experimental data, marked by Exp (see text). 
Our average results (LJ and LJ$^*$) are those from Table~\ref{TabTheoComp} and
the results marked by LJ$^*_{\rm E}$ are those obtained from LJ$^*$ by forcing
the result for $a_{+,0}$ to agree with the experimental data within 10\% (see text).
The corresponding relative errors are indicated between square brackets.}}
\begin{ruledtabular}
\begin{tabular}{lcccc}
 & ${a}_{+,0}/r_{\rm vdW}$ & ${a}_{+,1}/r_{\rm vdW}$ & ${a}_{+,1}/{a}_{+,0}$ & ${a}_{*,1}/r_{\rm vdW}$ \\
\hline
Exp & 2.50[0.10] & 39.3[0.12] & 15.7[0.22] & 4.78[0.20]\\
ZR        & 2.05[0.22] & 46.5[0.16] & 22.7[0.31] & 10.4[0.54]\\
LJ         & 1.41[0.77] & 27.7[0.42] & 19.7[0.20] & 3.33[0.43]\\
LJ$^*$ & 1.63[0.53] & 32.0[0.23] &19.7[0.20] & 3.85[0.24]\\
LJ$^*_{\rm E}$ & 2.27[0.10] & 44.7[0.12] &19.7[0.20] & 5.37[0.11]\\
\end{tabular}
\end{ruledtabular}
\label{TabExpComp}
\end{table}

{Evidently, there is much to be understood on the effects that realistic interactions can impose
in the determination of the three-body parameters. 
In more realistic systems the short-range multichannel nature of the interactions can produce, for instance, a 
different mixing of $s$- and $d$-wave components than the single channel model does. 
One can expect $d$-wave interactions to be more important when the system possesses a small background scattering 
length, i.e., of the order of $r_{\rm vdW}$, since in this case the entrance channel physics, obeying the universality of 
the van der Waals interactions \cite{gao2000pra}, can include a weakly bound $d$-wave state. 
Finite-width effects can lead to values of the effective range different than the one produced in our model, 
also determined by the universal van der Waals physics \cite{flambaum1999pra}. Such effects, although not entirely understood
yet, can also lead to substantial deviations of the three-body parameters \cite{schmidt2012epjb}. 
In fact, the model developed in Ref.~\cite{wang2014np}, which incorporates some 
of the multichannel physics of the problem, shows a much better agreement between theory and experiment \cite{Zenesini2014}, 
including for the $a<0$ geometric scaling $a_{-,1}/a_{-,0}\approx21.0$ from Ref.~\cite{huang2014prl},
indicating that both $s$- and $d$-wave mixing and finite-width effects might be at the heart of deviations of the three-body parameters 
for $a>0$ here obtained, as well as the deviations among the currently available experimental data (Table~\ref{TabExp}).
A fundamental difference between the physics for $a<0$ (where a more robust universal picture was found \cite{wang2012prl,dincao2013fbs,wang2012prlb,schmidt2012epjb,naidon2014pra,naidon2014prl,blume2015fbs} ---see Refs.~\cite{wang2013aamop,Ferlaino2011} 
for a summary of such experimental findings) and for $a>0$ is that corrections for the energy of the weakly bound $s$-wave dimer,
whether originated from mixing of $s$- and $d$-wave interactions or finite-width effects, should already 
lead to modifications on the $a>0$ three-body parameters. For $a_+$, the atom-dimer channel controls the interference effects
in recombination via the exit channel while it represents the initial collision channel responsible for the resonant effects 
determining $a_*$. In fact, under this perspective, a simple criteria can be established to determine whether 
$s$- and $d$-wave mixing and finite-width effects are important: if the degree of deviation between the binding energy obtained from multichannel 
interactions and the one obtained from single channel models are substantially different, such effects are likely to be
important.}

\section{Summary}

In conclusion, our present study establishes the universal values for the three-body parameters $a_*$ and $a_+$, both relevant
for ultracold quantum gases with {positive scattering lengths}. 
One of the most interesting results that has emerged from this study is the fact that the first excited Efimov resonance fails to intersect 
the dimer threshold, which is a surprising difference from the zero-range universal theories that always predict such an intersection.  
Our interpretation, that this failure of the resonance to intersect the threshold derives from important $d$-wave interactions, 
is consistent with findings from another recent study of this $a>0$ region \cite{giannakeas2016arXiv}
which uses a nonlocal potential model having no $d$-wave physics, and which {\it does} show such an intersection. 
The robustness of the present prediction thus hinges critically on whether the $d$-wave two-body physics is tightly constrained in 
the way predicted by van der Waals physics in single channel potential models \cite{gao2000pra,wang2012pra}. 
{Whether it is reasonable to expect that in the case of broad two-body Fano-Feshbach 
resonances, this linkage of two-body $s$-wave and $d$-wave resonance positions is satisfied, remains an open question deserving 
further investigation.}
However, especially in the case of narrow 
two-body resonances, $s$-wave and $d$-wave resonances are likely to be largely uncorrelated which presumably invalidates the present 
predictions in the vicinity of $a/r_{\rm vdW} \approx1$.
Nevertheless, the qualitative agreement between our results and the currently available experimental data partially 
confirms the notion of universality of Efimov physics for ultracold atoms. However, more experimental data and more sophisticated
theoretical models incorporating the multichannel nature of the atomic interactions might be necessary in order to quantitatively address 
present discrepancies.

\acknowledgments

This work was supported in part by the U.S. National Science Foundation (NSF) Grants PHY-1607204 and PHY-1607180
and by the National Aeronautics and Space Administration (NASA). PMAM also acknowledges JILA for the hospitality during his 
stay. The authors acknowledge S. Kokkelmans and P. Giannakeas for fruitful discussions.


\begin{thebibliography}{99}

\bibitem{Efimov1970} V. Efimov, 
Yad. Fiz. {\bf 12}, 1080 (1970); Sov. J. Nucl. Phys. {\bf 12}, 589 (1971).

\bibitem{braaten2006pr} E. Braaten and H.-W. Hammer, 
Phys. Rep. {\bf 428}, 259 (2006).

\bibitem{wang2013aamop} Y. Wang, J. P. D'Incao, and B. D. Esry, 
Adv. At. Mol. Opt. Phys. {\bf 62}, 1 (2013).

\bibitem{fletcher2013prl} R. J. Fletcher, A. L. Gaunt, N. Navon, R. P. Smith, and Z. Hadzibabic,
Phys. Rev. Lett. {\bf 111}, 125303 (2013).

\bibitem{rem2013prl} B. S. Rem, A. T. Grier, I. Ferrier-Barbut, U. Eismann, T. Langen, N. Navon, L. Khaykovich, F. Werner, D. S. Petrov, F. Chevy, and C. Salomon,
Phys. Rev. Lett. {\bf 110}, 163202 (2013).

\bibitem{makotyn2014np} P. Makotyn, C. E. Klauss, D. L. Goldberger, E. A. Cornell, and D. S. Jin, 
Nat. Phys. {\bf 10}, 116 (2014).

\bibitem{fletcher2016arxiv} R. J. Fletcher, R. Lopes, J. Man, N. Navon, R. P. Smith, M. W. Zwierlein, and Z. Hadzibabic,
 arXiv:1608.04377 (2016).
 
\bibitem{sykes2014pra}
A. G. Sykes, J. P. Corson, J. P. D'Incao, A. P. Koller, C. H. Greene, A. M. Rey, K. R. A. Hazzard, and J. L. Bohn
Phys. Rev. A {\bf 89}, 021601(R) (2014).

\bibitem{laurent2014prl} S. Laurent, X. Leyronas, and F. Chevy,
Phys. Rev. Lett. {\bf 113}, 220601 (2014).

\bibitem{smith2014prl} D. H. Smith, E. Braaten, D. Kang, and L. Platter,
Phys. Rev. Lett. {\bf 112}, 110402 (2014).

\bibitem{piatecki2014nc} S. Piatecki and W. Krauth,
Nat. Comm. {\bf 5}, 3503 (2014).

\bibitem{barth2015pra} M. Barth and J. Hofmann,
Phys. Rev. A {\bf 92}, 062716 (2015).

\bibitem{chin2010rmp} C. Chin, R. Grimm, P. S. Julienne, and E. Tiesinga,
Rev. Mod. Phys. {\bf 82}, 1225 (2010).

\bibitem{berninger2011prl} M. Berninger, A. Zenesini, B. Huang, W. Harm, H.-C. N\"{a}gerl, F. Ferlaino, R. Grimm, P. S. Julienne, and J. M. Hutson, 
Phys. Rev. Lett. {\bf 107}, 120401 (2011).

\bibitem{wang2012prl} J. Wang, J. P. D'Incao, B. D. Esry, and C. H. Greene, 
Phys. Rev. Lett. {\bf 108}, 263001 (2012).

\bibitem{dincao2013fbs} J. P. D'Incao, J. Wang, B. D. Esry, and C. H. Greene, 
Few-Body Systems {\bf 54}, 1523 (2013).
                
\bibitem{wang2012prlb} Y. Wang, J. Wang, J. P. D'Incao, and C. H. Greene,
Phys. Rev. Lett. {\bf 109}, 243201 (2012); {\em ibid.} {\bf 115}, 069901 (2015).

\bibitem{schmidt2012epjb} R. Schmidt, S.P. Rath, and W. Zwerger,
Eur. Phys. J. B {\bf 85}, 386 (2012).

\bibitem{naidon2014pra} P. Naidon, S. Endo, and M. Ueda,
Phys. Rev. A {\bf 90}, 022106 (2014).

\bibitem{naidon2014prl} P. Naidon, S. Endo, and M. Ueda,
Phys. Rev. Lett. {\bf 112}, 105301 (2014).

\bibitem{blume2015fbs} D. Blume, 
Few-Body Syst. {\bf 56}, 859 (2015).

\bibitem{dincao2005prl} J. P. D'Incao and B. D. Esry,
Phys. Rev. Lett. {\bf 94}, 213201 (2005).

\bibitem{braaten2007pra} E. Braaten and H.-W. Hammer,
Phys. Rev. A {\bf 75}, 052710 (2007).

\bibitem{nielsen1999prl} E. Nielsen and J. H. Macek, 
Phys. Rev. Lett. {\bf 83}, 1566 (1999).

\bibitem{esry1999prl} B. D. Esry, C. H. Greene, and J. P. Burke, 
Phys. Rev. Lett. {\bf 83}, 1751 (1999).

\bibitem{bedaque2000prl} P. F. Bedaque, Eric Braaten, and H.-W. Hammer,
Phys. Rev. Lett. {\bf 85}, 908 (2000).

\bibitem{bruch1973prl} L. W. Bruch and K. Sawada, 
Phys. Rev. Lett. {\bf 30}, 25 (1973).

\bibitem{lee2007pra} M. D. Lee, T. K\"ohler, and P. S. Julienne,
Phys. Rev. A {\bf 76}, 012720 (2007).

\bibitem{gao2000pra} Bo Gao,
Phys. Rev. A {\bf 62}, 050702(R) (2000).

\bibitem{wang2012pra} J. Wang, J. P. D'Incao, Y. Wang, and C. H. Greene,
Phys. Rev. A {\bf 86}, 062511 (2012).

\bibitem{wang2011pra} J. Wang, J. P. D'Incao, and C. H. Greene,
Phys. Rev. A {\bf 84}, 052721 (2011).

\bibitem{nielsen2002pra} E. Nielsen, H. Suno, and B. D. Esry, 
Phys. Rev. A {\bf 66}, 012705 (2002).


\bibitem{comment0} Based on the analysis of the hyperspherical adiabatic potentials we can trace
a strong coupling between the relevant $s$-wave channel for Efimov physics and the $d$-wave channel
associated with the $d$-wave dimer state, supporting a universal three-body state \cite{wang2012pra}

\bibitem{dincao2008pra} J. P. D'Incao, B. D. Esry, and C. H. Greene,
Phys. Rev. A {\bf 77}, 052709 (2008).

\bibitem{hutson2007njp} J. M. Hutson, 
New J. Phys. {\bf 9}, 152 (2007).

\bibitem{suno2002pra} 
H. Suno, B. D. Esry, C. H. Greene, and J. P. Burke, Phys. Rev. A {\bf 65}, 042725 (2002).

\bibitem{dincao2004prl}  
J. P. D'Incao, H. Suno, and B. D. Esry, Phys. Rev. Lett. {\bf 93}, 123201 (2004).

\bibitem{comment1} The results from Ref.~\cite{braaten2006pr} were multiplied
by a factor $3\sqrt{3}$ in order to ensure the proper comparison.

\bibitem{gogolin2008prl} A. O. Gogolin, C. Mora, and R. Egger, 
Phys. Rev. Lett. {\bf 100}, 140404 (2008).

\bibitem{deltuva2012pra} A. Deltuva, 
Phys. Rev. A {\bf 85}, 012708 (2012).

\bibitem{kievsky2013pra} A. Kievsky and M. Gattobigio,
Phys. Rev. A {\bf 87}, 052719 (2013).

\bibitem{ji2012ap} C. Ji, D. R. Phillips, and L. Platter,  
Ann. Phys. (NY) {\bf 327}, 1803 (2012).

\bibitem{platter2009pra} L. Platter, C. Ji, and D. R. Phillips, 
Phys. Rev. A {\bf 79}, 022702 (2009).

\bibitem{ji2015pra} C. Ji, E. Braaten, D. R. Phillips, and L. Platter,
Phys. Rev. A {\bf 92}, 030702(R) (2015).

\bibitem{li2016pra} J.-L. Li, X.-J. Hu, Y.-C. Han, and S.-L. Cong, Phys. Rev. A {\bf 94}, 032705 (2016).

\bibitem{Kraemer2006} T. Kraemer, M. Mark, P. Waldburger, J. G. Danzl, C. Chin, B. Engeser, A. D. Lange, K. Pilch, A. Jaakkola, H.-C. N\"{a}gerl and R. Grimm, 
Nature {\bf 440}, 315 (2006).

\bibitem{Ferlaino2011} F. Ferlaino, A. Zenesini,M. Berninger, B. Huang, H.-C. N\"{a}gerl, and R. Grimm, 
Few-Body Syst. {\bf 51}, 113 (2011).

\bibitem{Pollack2009} S. E. Pollack, D. Dries, and R. G. Hulet, 
Science {\bf 326}, 1683 (2009).

\bibitem{Dyke2013} P. Dyke, S. E. Pollack, and R. G. Hulet, 
Phys. Rev. A {\bf 88}, 023625 (2013).

\bibitem{Gross2009} N. Gross, Z. Shotan, S. Kokkelmans, and L. Khaykovich, 
Phys. Rev. Lett. {\bf 103}, 163202 (2009).

\bibitem{Gross2011} N. Gross, Z. Shotan, O. Machtey, S. Kokkelmans, and L. Khaykovich, 
Comp. Rend. Phys. {\bf 12}, 4 (2011).

\bibitem{Gross2010} N. Gross, Z. Shotan, S. Kokkelmans, and L. Khaykovich, 
Phys. Rev. Lett. {\bf 105}, 103203 (2010).

\bibitem{Zaccanti2009} M. Zaccanti, B. Deissler, C. D'Errico, M. Fattori, M. Jona-Lasinio, S. M\"{u}ller, G. Roati, M. Inguscio, and G. Modugno, 
Nat. Phys. {\bf 5}, 586 (2009). 

\bibitem{Knoop2009} S. Knoop, F. Ferlaino, M. Mark, M. Berninger, H. Sch\"{o}bel,
H.-C. N\"{a}gerl, and R. Grimm, Nat. Phys. {\bf 5}, 227 (2009).

\bibitem{Zenesini2014} A. Zenesini, B. Huang, M. Berninger, H.-C. N\"{a}gerl, F. Ferlaino, and R. Grimm, 
Phys. Rev. A {\bf 90}, 022704 (2014).

\bibitem{Machtey2012a} O. Machtey, Z. Shotan, N. Gross, and L. Khaykovich, 
Phys. Rev. Lett. {\bf 108}, 210406 (2012).

\bibitem{Lompe2010a} T. Lompe, T. B. Ottenstein, F. Serwane, K. Viering, A. N. Wenz, G. Z\"{u}rn, and S. Jochim, 
Phys. Rev. Lett. {\bf 105}, 103201 (2010).

\bibitem{roy2013prl} S. Roy, M. Landini, A. Trenkwalder, G. Semeghini, G. Spagnolli, A. Simoni, M. Fattori, M. Inguscio, and G. Modugno,
Phys. Rev. Lett. {\bf 111}, 053202 (2013).

\bibitem{Machtey2012b} O. Machtey, D. A. Kessler, and L. Khaykovich, Phys. Rev. Lett. {\bf 108}, 130403 (2012).

\bibitem{Langmack2012} C. Langmack, D. H. Smith, and E. Braaten, Phys. Rev. A {\bf 86}, 022718 (2012).

\bibitem{Hu2014} M.-G. Hu, R. S. Bloom, D. S. Jin, and J.M. Goldwin, Phys. Rev.
A {\bf 90}, 013619 (2014).

\bibitem{flambaum1999pra} V. V. Flambaum, G. F. Gribakin, and C. Harabati,
Phys. Rev. A {\bf 59}, 1998 (1999).

\bibitem{wang2014np} Y. Wang and P. S. Julienne,
Nat. Phys. {\bf 10}, 768 (2014).

\bibitem{huang2014prl} B. Huang , L. A. Sidorenkov, R. Grimm, and J. M. Hutson, 
Phys. Rev. Lett. {\bf 112}, 190401 (2014).

\bibitem{giannakeas2016arXiv} P. Giannakeas and C. H. Greene, arXiv:1608.08276 (2016).

\end{thebibliography}
\end{document}